# Building a Decision Tree Model for Academic Advising Affairs Based on the Algorithm C4. 5

**Mohammed Al-Sarem**

**Information Science Department, Taibah University**
**Al-Madinah Al-Monawarah, Kingdom of Saudi Arabia**

## Abstract

The ability to recognize students' weakness and solve any problem that may confront them in timely fashion is always a target for all educational institutions. Thus, colleges and universities implement the so-called academic advising affairs. On the academic advisor relies the responsibility of solving any problem that may confront students' learning progress.

This paper shows how the advisor can benefit from data mining techniques, namely decision trees techniques. The C4.5 algorithm is used as a method for building such trees. The output is evaluated based on the accuracy measure, Kappa measure, and ROC area. The difference between the registered and gained credit hours is considered as the main attribute on which advisor can rely.

## 1. Introduction

Academic advisory is known as a "process in which advisor and advisee enter a dynamic relationship respectful of the student's concerns" [1]. Often, this relationship relies on personal interactions between advisor (often an academic member) and advisee (students). The advisor's role is to draw the academic, social or personal directions to college student. Such directions might take the form of information, suggestion, counsel, discipline, coach, mentoring, or even teaching [2]. For this end, colleges and universities began to implement the so-called academic advising affairs. The academic advisory process, in several universities, is still manual. Henning in [3] discussed the problems of the manual advising process such as limited number of advisors, advisors availability, the problem of incompetent advisors, as well as, the serious consequences that may occur if mistakes are made, like; graduation delay, and major or college drop out. Even if the number of advisors is enough, this process is still a time-consuming task.

To cope with this problem and the problem of incompetent advisors, different applications have been used. Mostafa et al., [5] proposed a CBR advising system that can be used for converting the manual process of academic advising into an automated one. The system recommends to the student the most suitable major in his case by checking the similarity between the student taken course and the stored course in each departments. Another attempt to build an advising environment is the rule-based advisory system proposed by Al Ahmar [6] that aims to support students, in particular students of Information Systems (IS) major, during the selection of their courses for each and provide them and their advisors with all possible alternatives. The proposed system is also embedded with a quick and easy course selection and evaluation tool. In [4], the authors proposed an advising system to help undergraduate students during the registration period. The proposed system uses a real data from the registration pool, then applies association rules algorithm to help both students and advisors in selecting and prioritizing courses.

Traditionally, these systems depend greatly on the effort of the advisor to find the best selection of courses to improve students' performance [4]. However, the role of the academic advisor is not limited to finding the best selection of courses, but also to detect those students who have learning barriers. For this purpose, this paper presents an approach based on data-mining methodologies namely, decision trees [7]. The decision trees can apply in several real-world applications as a powerful solution to classification problem [8]. From an advisory point of view, the advisor looks for those students who have problems (or are closer to have problems) by mapping a sample into one predefined class. The classification process starts after giving as input sets of samples that consist of vectors of attribute values and a corresponding class [9]. A simple academic advisory classification might group students into three groups based on their performance: (1) those students whose grade point average GPA are below 2 (out of 5) (2) those students whose GPA are between 2 and 2.75 and (3) those students whose GPA are above 2.75.

## 2. Decision Trees

Decision trees are well-known methods for assisting the decision making process[10]. They can be considered as a directed graph where the root node has no incoming edges and other nodes have only one incoming edge. The construction process starts by splitting the instance space into several parts until finding so-called leaf node. The leaf node forms a decision rule which is used to determine the class of a new instance.

One of the well-known decision tree algorithms is C4.5 [10]. The C4.5 is a standard algorithm for inducing classification rules in the form of decision tree [11]. The *information gain* and *information gain ratio* are the criteria for choosing splitting attributes. The *information gain* is formalized as follows [12]:

$$gain = info(T) - \sum_{i=1}^{S} \frac{|T_i|}{|T|} \times info(T_i) \quad (1)$$

Where:

$$info(T) = -\sum_{j=1}^{N\ Class} \frac{freq(C_j, T)}{|T|} \times \log_2 \frac{freq(C_j, T)}{|T|} \quad (2)$$

is the entropy function. The C4.5 algorithm considers the information gain ratio of the splitting $T_i$ which is the ratio of $gain$ to its split information [10]:

$$Split(T) = -\sum_{i=1}^{S} \frac{|T_i|}{|T|} \times \log_2 \frac{|T_i|}{|T|} \quad (3)$$

Table 1: Pseudo-code of the C4.5 Algorithm [12]

| *FromTree (T)* |
|---|
| (1)  ComputeClassFrequency ($T$); |
| (2)  If OneClass or FewCases |
|     return a leaf; |
|     Create a decision node $N$ |
| (3)  ForEach Attribute $A$ |
|     ComputeGain(A); |
| (4)  $N.test$ = AttributeWithBestGain; |
| (5)  If $N.test$ is continous |
|     find Threshold |
| (6)  ForEach $T'$ in the splitting of $T$ |
| (7)    If $T'$ is Empty |
|       Child of $N$ is a leaf |
|     Else |
| (8)    Child of N= FromTree($T'$) |
| (9)    ComputeErrors of $N$; |
|     return $N$ |

## 3. Available data

The data available for research was collected from undergraduate students of Information Science department at Taibah University (TU), Al-Madinah Al-Munawarah, Kingdom of Saudi Arabia, using the existing academic system. The collected data was of three academic programs: old, new and developed. All available attributes are shown in Table II. The "*SId*" attribute, is included, here, in order to help the advisor in determining which academic plan of study was followed by student "*Plan_Study*". However, "*Ad_STATUS*" was determined by experienced advisor. The "*Total_Reg_C_H*" refers to the total registered hours in the system even if student decides to withdraw or postpone a course from the current semester after registration. The "*Diff_G_R_C_H*" shows the difference between the registered and gained hours. It is calculated by:

$$\Delta = \sum_{1}^{C} R(C_i) - \sum_{1}^{C} G(C_i) \quad (4)$$

Where:

$\Delta$- is different between the registered and gained credit hours.

$R(C_i)$ - is total credit hours for the registered courses of a student.

$G(C_i)$ - is total (actual) credit hours for the courses already taken and passed by a student.

The Grade Point Average a semester "*Sem_GPA*" is calculated as follows:

$$GPA_{Semester} = \frac{\sum_{i=1}^{C} w_i \times Credit\_Hour(C_i)}{\sum_{i=1}^{S} Credit\_Hour(C_i)} \quad (5)$$

Where:

$GPA_{Semester}$- is the Grade Point Average of a semester $S$.

$Credit\_Hour(C_i)$ - is the credit hours of a course $C_i$.

$w_i$- is the weight of the gained grade of a course $C_i$.

Table 2: Available data for classification task

| Attribute | Abbreviation | Attribute type | Availability By |
|---|---|---|---|
| Student Id | SId | | |
| Total Registered Credit Hours | Total_Reg_C_H | | |
| Total Gained Credit Hours | Total_Gain_C_H | | System |
| Total Credit Hours. in the Current Semester | Total_Cur_C_H | numeric | |
| Grade Point Average each Semester | Sem_GPA | | |
| Cumulated Grade Point Average | CUM_GPA | | Calculated by System |
| Different between Gained and Registered Credit hours | Diff_G_R_C_H | | |
| Category | Catg. | | |
| Learning Status | L_STATUS | | System |
| Gander | GEN | nominal | |
| Advisory status | Ad_STATUS | | |
| Academic Plan of Study | Plan_Study | | Advisor |

## 4. Classification Model

To help advisor in making decision about student's status, we find C4.5 algorithm is more suitable since the output results are shown graphically as decision tree. The Weka1 3.0 environment was used as a tool. The reason for this selection is that the Weka is the oldest and most successful open source data mining library[13]. To facilitate our work, we used the knowledge flow interface of Weka environment as alternative to the Explorer. The proposed model is shown in Fig.1.

In order to acquire good results, the "Sid", "GEN", "Sem_GPA" and "CUM. GPA" attributes were eliminated. The reason behind this elimination is that such attributes have not meaning, at least in our case, and both "Sem_GPA" and "CUM. GPA" attributes are correlated. To specify the attribute used by the class, the "Ad_STATUS" attribute is assigned as classification target class. For this purpose, the ClassAssigner tool was used. The 10-fold cross-validation also was applied on the C4.5 algorithm. The CrossValidationFoldMaker was attached

1 http://www.cs.waikato.ac.nz/ml/weka/downloading.html

and linked to the J48 component twice through the test set and training set options.

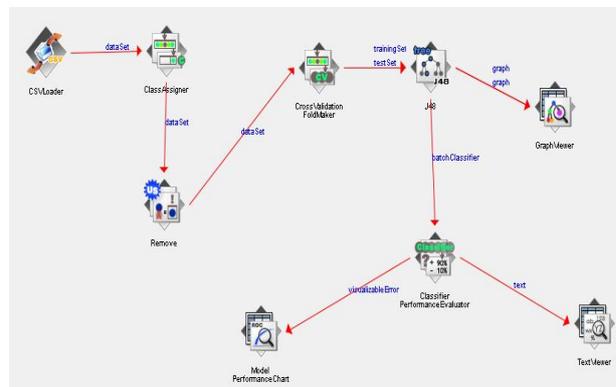

Figure 1: The proposed Classification Model

Finally, the Classifier PerformanceEvaluator tool was linked to J48 component via a batchClassifier. To visualize results, three visualization components: Graph viewer, Text viewer and Model PerformanceChart were used. The accuracy of the classifier is defined in terms of percentage of correct classified instances. Table3 and Table4 show the performance measures after applying the C4.5 algorithm.

Table 3:Performance measures' results

| correctly classified instances % | Kappa statistic | Mean absolute error | Root mean squared error | Relative absolute error % | Root relative squared error % |
|---|---|---|---|---|---|
| 87.550 % | 0.5461 | 0.1277 | 0.2751 | 61.31 % | 85.8932 % |

Since accuracy has disadvantages regarding a performance estimate and sensitivity to class distribution, other measures can take a place. To cope with these problems, the Kappa statistic [14] or Receiver Operating Characteristic (ROC) area [15] are good ways. The Kappa measures the agreement between two raters taking into account the agreement occurring by chance, whilst the ROC is used for visualizing, organizing and selecting classifiers based on their performance[16, 17].

Table 4: Detailed Accuracy

| TP Rate | FP Rate | Precision | Recall | F-Measure | ROC area | Class |
|---|---|---|---|---|---|---|
| 0.956 | 0.467 | 0.903 | 0.956 | 0.929 | 0.734 | Normal |
| 0.4 | 0.047 | 0.583 | 0.4 | 0.475 | 0.674 | Near To Risk |
| 0.900 | 0.000 | 1.000 | 0.90 | 0.947 | 0.944 | In Risk |
| 0.876 | 0.389 | 0.862 | 0.876 | 0.866 | 0.734 | Weighted Avg. |

From table 4, it is apparent that students whose performance is under the acceptable level (In risk class) were classified more precisely by the C4.5 algorithm.

However, the algorithm preformed quite well regarding the other two categories.

## 5. The C4.5 Decision Tree

The decision tree resulted from the C4.5 algorithm produces some interesting rules. These rules can be presented in both text and graph format. Figure 2 shows the most interested output of C4.5 algorithm in the text format.

```
Learning Status = In Study
| Different between Gained and Registered Credit hour <= 36: Normal (180.0/19.0)
| Different between Gained and Registered Credit hour > 36
| | Total Registered Credit Hour <= 137: Near To Risk (8.0)
| | Total Registered Credit Hour > 137
| | | Total Registered Credit Hour <= 157: Normal (6.0/1.0)
| | | Total Registered Credit Hour > 157: Near To Risk (5.0)
```

Figure 2: the C4.5 interested outputs

Below, examples of interpretation of the decision tree's branches:

> If student is still studying and the difference between the registered credit hours and the gained credit hours "*Diff_G_R_C_H*" is less than 36 credit hours, then he/she is likely to finish his/her study without any problems.

> If student is still studying and the difference "*Diff_G_R_C_H*" between the registered credit hour and the gained credit hour" is greater than 36 credit hours and the Total Registered Credit Hours "*Total_Reg_C_H*" is less than 137 or greater than 157, then the student is likely to have problems and the advisor should be careful in selecting courses.

> If student is still studying and the difference "*Diff_G_R_C_H*" between the registered credit hour and the gained credit hour" is greater than 36 credit hours and the Total Registered Credit Hours "*Total_Reg_C_H*" is between 137 and 157, then the student is likely to finish his/her study without any problems.

## 6. Conclusion and Future Work

The current research showed how we can use the C4.5 algorithm in supporting academic advisors during their work. The accuracy of classifier, Kappa coefficient and

ROC area were used as measures tool to evaluate algorithm's output. The results showed some interesting rules. The difference between the registered and gained credit hours by a student was the main attribute that academic advisors can rely on.

In the future, we are planning to apply other data mining algorithms using additional data about the students of other departments. We also intend to find other interesting rules that can be helpful for academic advisory.

### Acknowledgments

I would like to thanks students' affair and registration faculty of Taibah University, KSA for facilitating access to students' academic data that were used during this research. also, I thank dr. Mohammed Muharram, the head of department of English and director of TOFEL center at Thamar University, Republic of Yemen for proofreading this work.

**Mohammed Al-Sarem:** Dr. Al-Sarem is an assistant professor of information science at the Taibah University, Al Madinah Al Monawarah, KSA. He received the PhD in Informatics from Hassan II University, Mohammadia, Morroco in 2014. His research interests center on E-learning, educational data mining, Arabic text mining, and intelligent and adaptive systems. He published several research papers and participated in several local/international conferences.
.